\documentclass[prb,twocolumn,superscriptaddress]{revtex4-2}
\usepackage{amsmath,amssymb,bm,mathrsfs,graphicx, braket, times,color}
\usepackage[colorlinks=true,citecolor=green!70!black]{hyperref}
\usepackage{soul}
\usepackage{xcolor}
\usepackage{colortbl}
\usepackage[makeroom]{cancel}
\usepackage{array,multirow,graphicx}
\usepackage{tikz}
\usetikzlibrary{shadings}
\usepackage{txfonts}  
\usepackage{txfontsb} 
\newcommand{\rv}{\mathbf r}
\newcommand{\uv}{\mathbf{u}}
\newcommand{\nablav}{\bm{\nabla}}


\begin{document}

\title{Nanomotion of micro-objects driven by light-induced elastic waves on solid interfaces}

\author{Wei Lyu}
\affiliation{Zhejiang University, Hangzhou 310027, Zhejiang Province, China}
\affiliation{Key Laboratory of 3D Micro/Nano Fabrication and Characterization of Zhejiang Province, School of Engineering, Westlake University, 18 Shilongshan Road, Hangzhou 310024, Zhejiang Province, China}
\affiliation{Institute of Advanced Technology, Westlake Institute for Advanced Study, 18 Shilongshan Road, Hangzhou 310024, Zhejiang Province, China}

\author{Weiwei Tang}
\affiliation{College of Physics and Optoelectronic Engineering, Hangzhou Institute for Advanced Study, University of Chinese Academy of Sciences, No. 1, Sub-Lane Xiangshan, Xihu District, Hangzhou 310024, China }

\author{Wei Yan}
\thanks{wyanzju@gmail.com}
\affiliation{Key Laboratory of 3D Micro/Nano Fabrication and Characterization of Zhejiang Province, School of Engineering, Westlake University, 18 Shilongshan Road, Hangzhou 310024, Zhejiang Province, China}
\affiliation{Institute of Advanced Technology, Westlake Institute for Advanced Study, 18 Shilongshan Road, Hangzhou 310024, Zhejiang Province, China}

\author{Min Qiu}
\thanks{qiumin@westlake.edu.cn }
\affiliation{Key Laboratory of 3D Micro/Nano Fabrication and Characterization of Zhejiang Province, School of Engineering, Westlake University, 18 Shilongshan Road, Hangzhou 310024, Zhejiang Province, China}
\affiliation{Institute of Advanced Technology, Westlake Institute for Advanced Study, 18 Shilongshan Road, Hangzhou 310024, Zhejiang Province, China}

\begin{abstract}
It has been recently reported that elastic waves induced by nanosecond light pulses can be used to drive nano-motion of micro-objects on frictional solid interfaces, a challenging task for traditional techniques using tiny optical force. In this technique, the main physical quantities/parameters involved are: temporal width and energy of light pulses, thermal heating and cooling time, friction force and elastic waves. Despite a few experimental observations based on micro-fiber systems, a microscopic theory, which reveals how these quantities collaboratively enable motion of the micro-objects and derives what the underlying manipulation principles emerge, is absent. In this article, a comprehensive theoretical analysis—centralized around the above listed physical quantities, and illuminated by a single-friction-point model in conjunction with numerical simulations—is established to pedagogically clarify the physics. Our results reveal the two essential factors in this technique: (1) the use of short light pulses for rapid thermal expansion overwhelming friction resistance and (2) the timescale asymmetry in thermal heating and cooling for accumulating a net sliding distance. Moreover, we examine the effects of spatially distributed friction beyond the single-friction-point consideration, and show “tug-of-war”-like friction stretching in the driving process. Given these insights, we positively predict that this elastic-wave-based manipulation principle could be directly translated to micro/nano-scale optical waveguides on optical chips, and propose a practical design. We wish that these results offer theoretical guidelines for ongoing efforts of optical manipulation on solid interfaces with light-induced elastic waves.
\end{abstract}
\maketitle

\section{Introduction}

Optical manipulation of micro- and nano-scaled objects provides substantial applications~\cite{Xin2020}, such as single cell analysis~\cite{Ashkin1987,Ramser2010,Zhong2013} and drug delivery~\cite{Kang2013,MacDonald2003} in microorganism biology, atomic quantum processor~\cite{Tamara2021,Ren2021a} in physics, and microfluidic devices~\cite{Awel2021}. According to their different mechanisms and principles, the implementation routes of optical manipulations can be generally divided into optical tweezers using optical forces~\cite{Ashkin1986, Grier2003, Bustamante2021} or dielectrophoresis-based-methods~\cite{Ren2021,Lin2020,Lin2018,Park2009,Zheng2022a}. In all these techniques, it is crucial to mitigate adhesion between driven objects and their environments. Consequently, it is a common practice to carry out optical manipulation in liquids~\cite{Terray2002,Awel2021,Venu2013,Enachi2016} or by levitating nano-particles in vacuum~\cite{Hebestreit2018, Conangla2018,Pu2021}, wherein the adhesion becomes negligible and smaller than pN---the typical order of magnitude of optical/photophoretic forces~\cite{Chen2011, Arya2021, Gao2017} exerted on micro-objects.

\begin{figure}[ht!]
\includegraphics[width=0.48\textwidth]{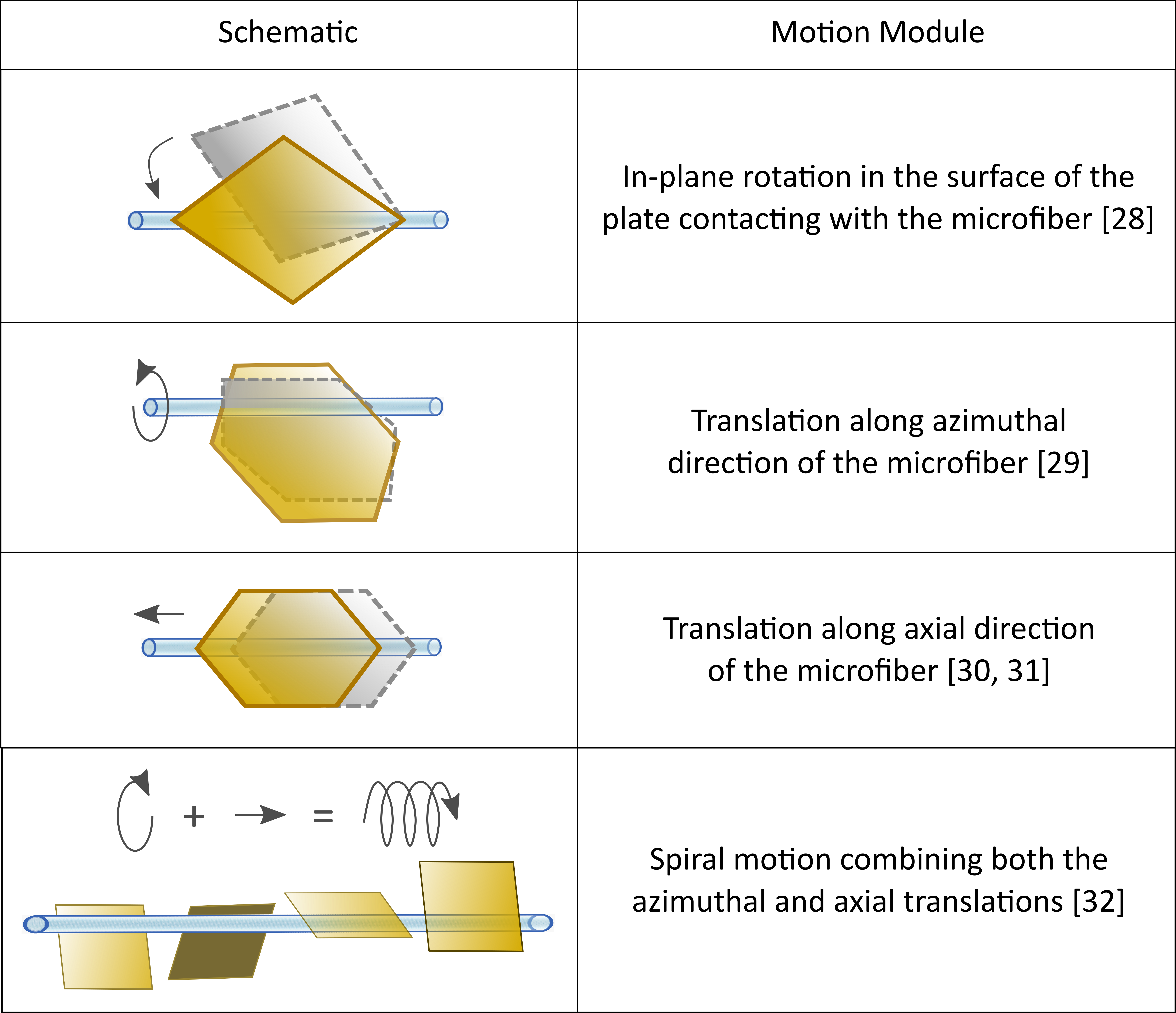}
\caption{ {\bf Existing experiments that report four motion modules of nano-motion of gold plates driven by elastic waves induced by nanosecond light pulses.}
}
\label{fig::table1}
\end{figure}

Different from conventional routines, the researchers have recently proposed to exploit elastic waves induced by pulsed light to manipulate micro-objects. This opens up an alternative way to accomplish controllable and precise manipulation of micro-objects on solid interfaces, wherein the surface adhesion could easily exceed over $\mu$N~\cite{Kendall1994}. The reported experiments, which are performed in micro-fiber-based systems using nano-second pulsed light, are briefly summarized in Fig.~\ref{fig::table1} for clarification. The motion is stepwise driven by single pulses, and, in each step, the motion distance is only a few nanometers or even sub-nanometers. Consequently, the motion speed could be controlled by varying repetition rates and energy of laser pulses. Notably, several distinct nano-motion modules of gold micro-plates have been demonstrated, including in-plane rotation in the surface of the plate contacting with the microfiber~\cite{Lyu2022}, translation along azimuthal~\cite{Lu2019} or axial~\cite{Lu2017,Gu2021} direction of the microfiber, and spiral motion combining both the azimuthal and axial translations~\cite{Tang2021}. These different nano-motion modules can be controlled by carefully adjusting  relative positions and contact configurations between the driven micro-plates and the micro-fibers, or by exploiting specific spatial profiles of optical absorption. For instance, the translation along the azimuthal direction of the microfiber requires the geometrical asymmetry in the two wings of the plate, and the motion is towards the short-wing side~\cite{Lu2019}. On the other hand, the translation along the axial direction of the microfiber requires either the asymmetry in the contact between the plate and the micro-fiber~\cite{Tang2021} or the profile asymmetry of the optical absorption profile~\cite{Gu2021}, which, respectively, result in that the plate translates axially towards the contact side or the intense-absorption side. The combination of the azimuthal and axial translations gives the spiral motion. Lastly, the in-plane rotation is a consequence of the asymmetry in the two wings of the plate and the gradient distribution of the optical absorption power along the touching line between the plate and the micro-fiber~\cite{Lyu2022}.

These reported experiments of the elastic-wave-based nano-motion might inspire future developments of miniature motors on solid interfaces~\cite{Zheng2022b}. As a next step, a meaningful direction is to extend this technique from its initial micro-fiber-based system, e.g., to on-chip integration~\cite{Hunt2007,Ding2012,Ito2020,Tardif2022,Zhang2021,Li2021}, therein bringing many unique applications. For instance, by manipulating motion modules and controlling precise positions of a micro-object in a waveguide network, one could potentially modulate light flow, and, thus, realize a mobile optical modulator/switch. In addition, by arranging relative positions of a group of micro-objects on a substrate, reconfigurable optical devices can be achieved~\cite{Zheng2019}. Plenty of experimental efforts are clearly demanded. Meanwhile, it is of equal importance to establish a comprehensive, rigorous understanding of the physical mechanism, and, further, to fairly access the features of this technique, as has been done for optical tweezers. This is the main purpose of this article.

The current knowledge of the elastic-wave-based nano-motion considers dynamic interactions between elastic waves—due to temperature rising through optical absorption—and their induced friction force (the parallel component of surface adhesive force). The friction force is identified to drive the motion, similar to human walking. However, the existing knowledge is mainly proposed for phenomenologically interpreting the experimental observations, while being inefficient in quantitative predictions. This is because that the motion is a complicated optical-thermal-elastic process, wherein multiple physical parameters/quantities, such as temporal width and energy of light pulses, thermal heating and cooling time, friction resistance and elastic waves, are involved and couple with each other. Therefore, a precise, comprehensive understanding should necessarily take all these factors into account and classify their individual roles, which, however, still remains absent.

In this article, we fill this gap by starting with a two-dimensional (2D) physical model, in which, a 2D plate on a substrate is driven by light pulses, as is shown in Fig.~\ref{fig::sketchphys}A. Despite being slightly distant from the realistic experiments (summarized in Fig.~\ref{fig::table1}), this designed physical model advantageously allows us to reveal the different roles of the key physical quantities through assited analytic analysis. Along this way, we reveal how temporal width and energy of light pulses, thermal heating and cooling time, contact friction and its distribution impact the motion, and, further, how one could practically play these diverse factors benefiting the nano-motion. Particularly, to investigate interactions between friction force and light-induced elastic waves, a single-friction-point model is developed. These results are summarized in Sections II-VI.

In Section VII, we provide perspectives for future developments of this technique and, particularly, discuss its practical realization on optical chips. We numerically evidence that a gold micro-plate can be driven on a Si$_3$N$_4$ nano-waveguide, where the friction force reaches tens of $\mu$N. Finally, Section VIII concludes the article.

\section{Physical picture}

The nano-motion of a micro-object driven by light-induced elastic waves hosts rich opto-thermo-elastic physics. A rigorous theoretical description should combine three sets of equations: Maxwell’s equations, heat conduction equation and elastic wave equation, from which meaningful quantifiers/relations—that specify, characterize the motion—might be derived. However, this approach is more suitable for numerical simulations, but difficult to bring analytic insights due to its mathematical complexity.

\begin{figure}[th!]
\includegraphics[width=0.4\textwidth]{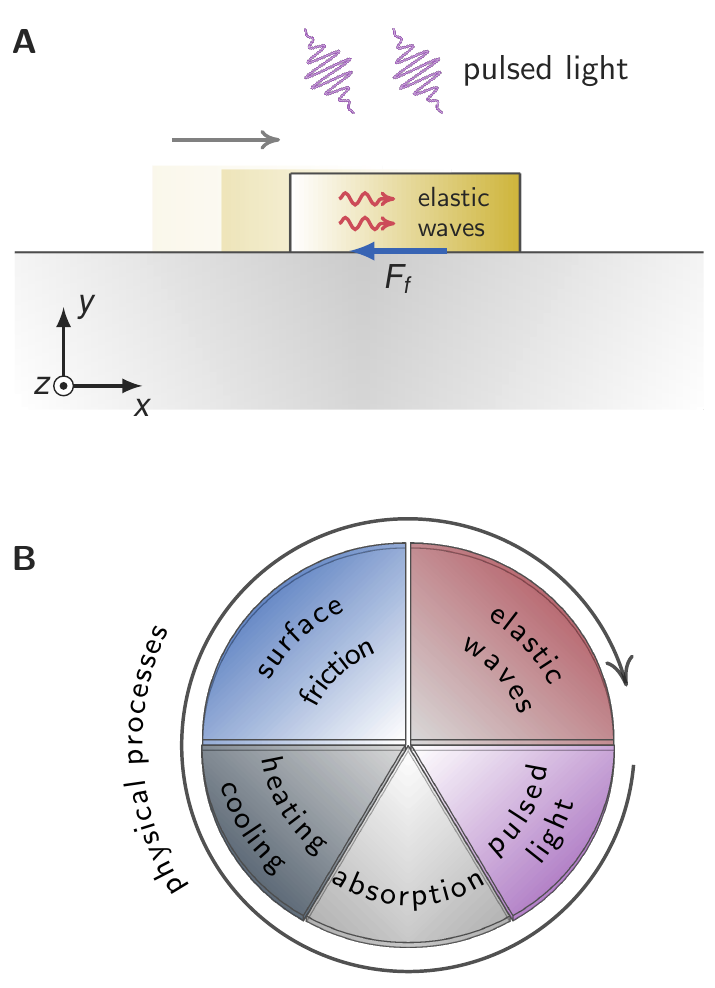}
\caption{{\bf (A) Sketch and (B) involved physical processes of nano-motion of a micro-object on a substrate driven by elastic waves induced by pulsed light. }
}
\label{fig::sketchphys}
\end{figure}

To bypass this technical difficulty, we omit less-important physical details, and instead focus directly on the central physics: light-induced thermal deformation in the presence of surface friction. To put our strategy concretely, we recall the involved physical processes (see Fig.~\ref{fig::sketchphys}B). First, under incidence of pulsed light, the micro-object converts light to heat. Then, the generated heat results in the temperature rising and excites elastic waves, therein leading to the thermal deformation of the micro-object. At the same time, the surface friction is induced to resist the deformation, and, counter-intuitively drives the motion of the micro-object. Apparently, the last step, concerning the interplay between the surface friction and the deformation, is the key in the whole processes, in which, the explicit role of the surface friction could be quantized by contrasting the deformation of the micro-object with and without the friction. Then, the involved physical quantities/parameters can be dependably discussed around this friction-induced deformation difference. Moreover, to better interpret the physics, we intentionally refer to a simple 2D model (see Fig.~\ref{fig::sketchphys}A), as has been mentioned in the introduction.

{\it Essential physical parameters}. --- For a better clarification of the physical dynamics in the aforementioned optical-thermal-elastic processes (see Fig.~\ref{fig::sketchphys}B), we introduce the following parameters:
\begin{itemize}
  \item $t_w$ and $W_{abs}$, temporal width and optical absorption energy of light pulses. In our study, light pulses are set to be Gaussian, and their temporal width $t_w$ is defined to be $1/e$ width of the pulse energy.
  \item $t_{heat}$ and $t_{cool}$, thermal heating and cooling time. With the use of short light pulses, $t_{heat}$ is roughly about $t_w$. $t_{cool}$ is defined as the time for the thermal energy decaying from peak to half.
  \item $F_f^s$, friction sliding resistance, which is the maximum allowable static friction exerted on the plate.
  \item $v_L$, velocity of excited elastic waves in the plate.
\end{itemize}

In addition to the above parameters, the other involved ones include: $t_R\equiv2L_p/v_L$, one-round-trip travel time of the elastic waves in the plate, where $L_p$ denotes the plate length in the $x$-direction (see Fig.~\ref{fig::sketchphys}A); $t_{diff}\equiv L_p^2/\alpha_{diff}$, characteristic time for heat spreading through the plate, where $\alpha_{diff}$ denotes the thermal diffusivity; $\lambda_{el}\equiv v_Lt_W$, characteristic wavelength of the excited elastic waves.

To have a concrete perception of these parameters, we estimate their magnitudes by considering a gold plate with $L_p=10\, \mu\rm m$ (typical length size of the micro-object used in the existing experiments of the nano-motion~\cite{Lu2017, Lu2019, Tang2021, Lyu2022}). For gold,  $v_L\sim\sqrt{E/\rho}$ ($E$, Young’s modulus; $\rho$, mass density) is about $2000\,\rm m/s$ and $\alpha_{diff}=\sqrt{k/\rho c_p} \simeq 10^{-4}\,\rm  m^2/s$ ($k$, thermal conductivity; $c_p$, specific heat capacity). Consequently, there are $t_R\simeq 10\, \rm ns$ and $t_{diff}\simeq 100\,\rm ns$. We thus observe that $t_R\ll t_{diff}$, that is, the elastic waves travel much faster than the heat diffusion. This relation implies that the object can be well deformed by the elastic waves at positions where the heat has not been yet spread. This observation leads us to the following deduction: the traditional picture of "static” thermal expansion and contraction, which is widely used for describing thermal deformation, is incomplete in the nanosecond regime, and the elastic waves should be taken into account.

$t_{cool}$, the cooling time, is mainly determined by the heat conduction from the plate into the substrate, which could vary from tens to thousands (or even more) of nanoseconds, depending on the thermal contact (see latter discussions concerning Fig.~\ref{fig::defwofric}). $t_W$, pulse duration, on the other hand, is an external parameter relating to the laser source used, which determines the heating time, $t_{heat}$. With $t_W$ in the order of nanoseconds, the characteristic wavelength of the excited elastic waves, $\lambda_{el}$, is estimated from its definition to be in the order of micrometers.

In the reported experiments where gold plates contact with micro-fibers with diameter of a few microns, the sliding resistance, $F_f^s$, is measured to be a few $\mu$N~\cite{Lu2019, Gu2021}. It is known that the $F_f^s$ is linearly proportional to the contact area. The contact area between the gold micro-plate and the micro-fiber with diameter of a few microns is estimated to be in the order of 10 $\mu$m$\times$10 nm, where 10 $\mu$m denotes the typical contact length along the axial direction of the micro-fiber, and 10 nm is the order of the transverse contact length that is calculated from the contact simulations considering the interfacial van der Waals forces (see Supplementary Fig. S1~\cite{SeeSupplementalMaterial}).

\section {Thermal deformation without friction in the nanosecond regime}

{\it General features}. --- We start by studying thermal deformation of a micro-object in the absence of surface friction under incidence of a single light pulse. The micro-object responses to the light-induced temperature change $\delta T$ and deforms its shape. The deformation is characterized by the coordinate changes of the structure, denoted by ${\uv}^{th} (\rv;t)$ (termed as elastic displacements hereafter), which satisfy~\cite{LandauBook}:
\begin{align}
\nablav\times \nablav \times \uv^{th} ({\rv};t) &-2\frac{1-\sigma}{1-2\sigma}\nablav\nablav\cdot  \uv^{th} ({\rv};t) + \nonumber \\
2\alpha_{th} \frac{1+\sigma}{1-2\sigma}\nablav\delta T (\rv;t)&+\frac{2\rho(1+\sigma)}{E}\frac{\partial^2\uv^{th}(\rv;t)}{\partial t^2}=0.
\label{eq:elaswave}
\end{align}
Here $\sigma$, $E$ and $\alpha_{th}$, respectively, denote Poisson’s ratio, Young’s modulus and the coefficient of linear thermal expansion of the micro-object. The temperature change $\delta T$ relates to the optical absorption. The temperature evolution undergoes heating phase during the short period of the pulse injection and the next cooling phase, during which the object successively expands and contracts its volume.

In the absence of the surface friction (interfacial force), the external force exerted on the micro-object is null. Consequently, although the elastic displacements $\uv^{th}$ exist, their spatial average (i.e., the centroid displacement) remains zero. The object is thus “motionless”. This can be examined by applying the volume integration to Eq. \eqref{eq:elaswave}, wherein its first three terms of the left side vanish, thus leaving  $\partial^2<\uv^{th} (t)>/\partial t^2=0$ with  $<{\uv}^{th}(t)>$ denoting the spatial average of $\uv^{th} (\rv;t)$.

With the use of (nano-second) pulsed light, as has been discussed in Sec. II, a complete description of the thermal deformation should take dynamic elastic-wave propagations into account, since $t_R\ll t_{diff}$. Referring to the studied 2D plate, the excited elastic waves are mainly carried by the fundamental longitudinal mode, with elastic displacements pointing along the propagation direction and uniformly distributing over the plate thickness. Specifically, the wavenumber $k_L$ and velocity $v_L$ of the fundamental longitudinal mode are given by~\cite{Tang2021}:
\begin{align}
k_L=\omega\sqrt{\rho(1-\sigma^2)/E},\quad  v_L=\sqrt{E/(\rho(1-\sigma^2))}.
\label{eq:elasK}
\end{align}
Note that Eq.~\eqref{eq:elasK} is valid under the assumption that $k_L h\ll 2\pi$ ($h$, plate thickness).

\begin{figure*}[ht!]
\includegraphics[width=0.82\textwidth]{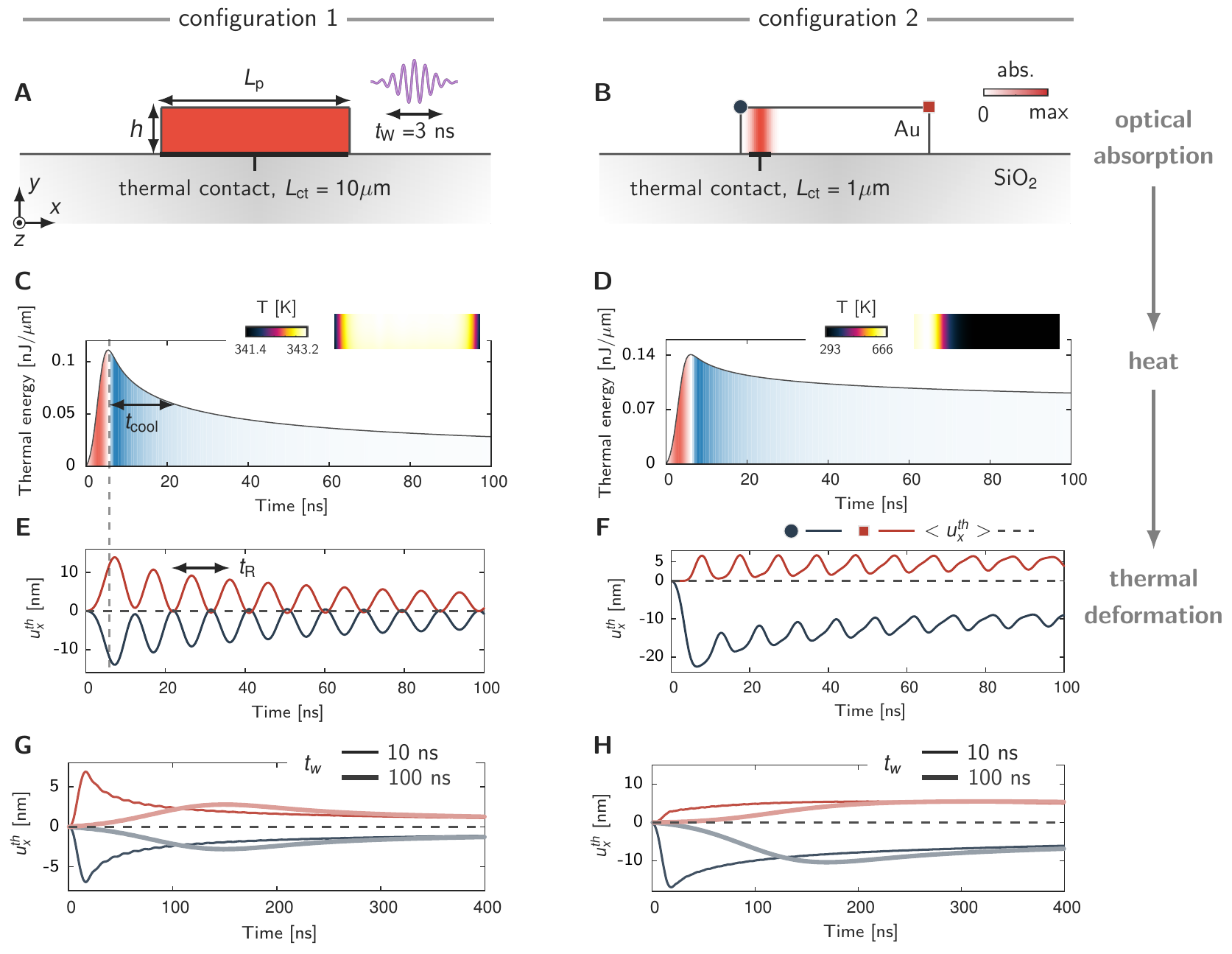}
\caption{{\bf Thermal deformation without friction driven by a nanosecond light pulse}.
{\bf A.-B.} Schematics of two studied configurations: ({\bf A}) uniform optical absorption in a gold micro-plate with a full thermal contact between the plate and the SiO$_2$ substrate; ({\bf B}) localized optical absorption with a partial thermal contact. The 2D plate in the $x-y$ plane has length and height of $L_p=10\, \mu \rm m$ and $h=50\, \rm nm$, respectively. In {\bf B}, the optical absorption has a Gaussian distribution in the $x$-direction, centered at 1 $\mu$m from the left edge of the plate, with $1/e$ width 1 $\mu$m; the thermal contact length is 1 $\mu$m with its center coinciding with that of the optical absorption.
{\bf C.-D.} Temporal evolutions of the thermal energy stored in the plate, highlighting the heating (red shaded areas) and cooling phases (blue shaded areas). Insets: temperature profiles of the plate at the peak of the thermal energy, $t=6\,\rm ns$. {\bf E.-F.} Temporal evolutions of (dominant) $x$-component elastic displacements induced by the temperature variation, $u_x^{th}$, of two edge points (red and blue lines; edge points are marked in {\bf B}), and of $x$-component centroid displacement, denoted by $<u_x^{th}>$ (dashed lines). In {\bf C-F}, the temporal width of the light pulse is $t_W=3\,\rm ns$ and the total optical absorption energy is $W_{abs}=0.2\, \rm nJ/\mu m$. Note the peak of the stored thermal energy is observed to be less than $W_{abs}=0.2\, \rm nJ/\mu\rm m$ due to the heat conduction from the plate into the substrate. The simulations are performed with COMSOL Multiphysics, and the material parameters are taken from the built-in MEMS library of COMSOL Multiphysics (see Supplementary Table 1~\cite{SeeSupplementalMaterial}). {\bf G.-H.} Same as {\bf E-F}, except that temporal widths of light pulses increases to $t_W=10, 100\, \rm ns$.
}
\label{fig::defwofric}
\end{figure*}

We numerically illustrate the general features of the light-pulse-induced thermal deformation in Fig.~\ref{fig::defwofric}. A gold 2D plate with thickness $h=50\,\rm nm$ and length $L_p=10 \,\mu\rm m$ is placed on a glass substrate. A Gaussian light pulse, with temporal width $t_W=3\,\rm ns$, is injected into the plate and leads to an absorption energy of $W_{abs}=0.2\,\rm nJ/\mu m$ (note that the unit of $W_{abs}$ is $\rm nJ/\mu m$ due to the consideration of the 2D plate).
We consider two representative configurations of the absorption distributions, (1) evenly and (2) locally distributed along the length direction (i.e., $x$-direction) of the plate, as is shown in Figs.~\ref{fig::defwofric}A and B, respectively, and their spatial distributions are described in the figure caption. Moreover, two different thermal contact scenarios are set to examine the effect of the cooling time $t_{cool}$ on the deformation. In the first case (Fig.~\ref{fig::defwofric}A), the plate and the substrate are assumed to be in full thermal contact, so that the thermal energy in the plate could be rapidly conducted into the substrate. In contrast, the other case (Fig.~\ref{fig::defwofric}B) models the slow cooling, where the plate and the substrate has a short thermal contact length of 1 $\mu$m (see the figure caption for more details). The thermal contact length practically relates to the surface topography of the plate~\cite{Tang2021}. The full contact is realizable when the plate is flat and the associated roughness is negligible, otherwise, the partial contact occurs.

{\it Heat evolution}. --- Figures~\ref{fig::defwofric}C and D plot the temporal evolutions of the thermal energy stored in the plate. Note the peak of the stored thermal energy is observed to be less than $W_{abs}=0.2\, \rm nJ/\mu m$ due to the heat conduction from the plate into the substrate. In the full-contact case (Fig.~\ref{fig::defwofric}C), the thermal energy quickly drops after peak due to the efficient heat conduction from the plate into the substrate. The cooling time $t_{cool}$ is estimated to be only about 10 ns. On the contrary, in the partial-contact case (Fig.~\ref{fig::defwofric}D), the heat conduction channel between the plate and the substrate is narrow, such that the cooling process exhibits a much longer tail with $t_{cool}$ exceeding over 100 ns.

The temperature distribution in the plate is the joint result of the optical absorption distribution and the heat diffusion. In the early period with  $t\ll t_{diff}$ ( $t_{diff}\simeq 100\, \rm ns$, time for heat fully diffusing the entire plate), the heat localizes around where the optical absorption takes place, and the temperature distribution resembles the absorption distribution, as shown in the insets of Figs.~\ref{fig::defwofric}C and D.

{\it Thermal deformation.} --- Figures~\ref{fig::defwofric}E and F trace the dominant $x$-component elastic displacements, $u_x^{th}$, of the left (dark blue lines) and right (dark red lines) ending points (marked in Fig.~\ref{fig::defwofric}B) of the plate. Note that, since we here consider a thin plate with thickness only about tens of nanometers, the deformation in the thickness direction is almost uniform. It is seen that $u_x^{th}$ oscillates with periodicity close to the one-round-trip travel time of the elastic waves ($t_R\simeq 10\,\rm ns$), which evidences the propagation of the elastic waves. Moreover, as indicated in Eq.\eqref{eq:elaswave}, the thermal deformation is directly determined by the temperature profile. In this regard, with a uniform temperature distribution, the left and right halves of the plate deform symmetrically, as verified in Fig.~\ref{fig::defwofric}E that $u_x^{th}$ of the left and right endpoints show same magnitude but opposite signs. On the contrary, if the temperature distribution is non-uniform, so is the deformation, i.e., featuring the asymmetric deformation, as is confirmed in Fig.~\ref{fig::defwofric}F.

During the period of the light pulse injection (the left regions bounded by the vertical dashed lines in Figs.~\ref{fig::defwofric}C and E), the thermal deformation rapidly intensifies to its maximum as the thermal energy climbs to the peak. Then, the thermal cooling initiates and the deformation gradually mitigates. The recovery rate of the deformation is determined by $t_{cool}$: a smaller value of $t_{cool}$ (i.e., faster thermal cooling) results in a faster deformation recovery and vice versa. Furthermore, regardless of the dynamic change of the local deformation, the plate centroid always remains zero (see the horizontal dashed curves in Figs.~\ref{fig::defwofric}E and F), since no external force is applied.

The temporal width of the light pulse $t_W$ also affects the thermal deformation. To illustrate this effect, we increase $t_W$  to 10 and 100 ns. The computed $x$-component elastic displacements, $u_x^{th}$, as functions of time, are plotted in Figs.~\ref{fig::defwofric}G and H. Markedly, as $t_W$ increases, the deformation oscillations become apparently weaker. This is because that the adjacent oscillations with the interval $t_R$ , due to the round trips of the elastic waves, are smeared out in a longer period of $t_W$ when the elastic waves are continuously excited. Given this intuition, we deduce that the visibility of the deformation oscillations requires that $t_W<t_R$ ($t_R\simeq 10\,\rm ns$ here), as is confirmed by comparisons between Figs.~\ref{fig::defwofric}E-F and G-H. Moreover, in the heating period, we find that the deformation speed ($\partial u_x^{th}/\partial t$) increases as $t_W$ decreases. However, in the cooling period, the rate of the deformation recovery is independent of $t_W$, and is instead determined by $t_{cool}$, as is seen in Figs.~\ref{fig::defwofric}E-H that the decaying tails of $u_x^{th}$ are similar for $t_W$ with different values.

\section{Accounting surface friction in thermal deformation: single-friction-point model}

\begin{figure}[th!]
\includegraphics[width=0.45\textwidth]{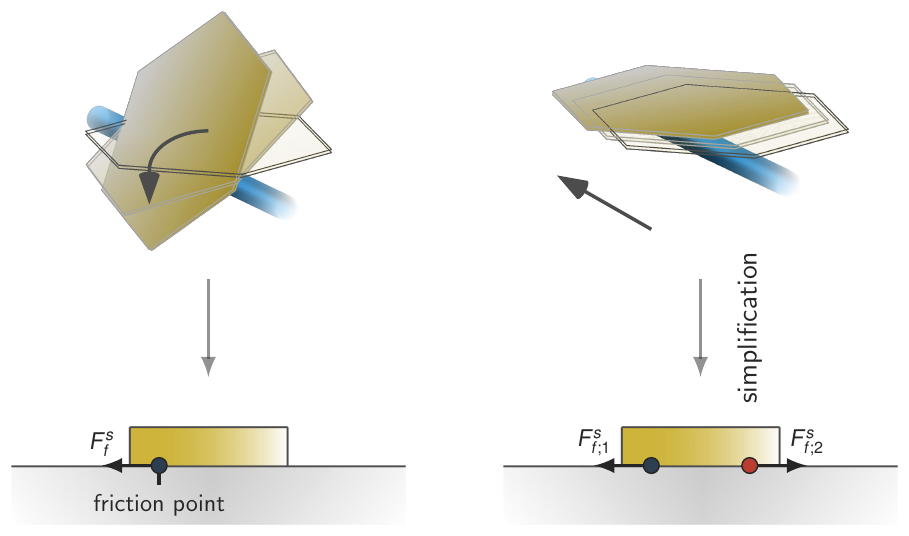}
\caption{{\bf Intuitive simplifications of complex contact scenarios to single friction point (left) and two friction points (right)}. The single-friction-point model well approximates the case where the contact length in the motion direction is negligible, such as that the azimuthal movement of a gold micro-plate around the micro-fiber~\cite{Lu2019}. On the other hand, the two friction points are introduced for the case when the contact length in the motion direction becomes noticeable comparing to the characteristic elastic wavelength or the object dimension, such as the axial movement of the plate along the micro-fiber~\cite{Lu2017,Gu2021}.
}
\label{fig::simple}
\end{figure}

{\it Introduction of single-friction-point model.} --- The friction force exerts on the plate when the latter slides on the substrate. To elucidate the effects of the friction on the thermal deformation, we establish a simplified physical model by assuming that the contact length along the sliding direction of the micro-object is significantly smaller than both the wavelength of the excited elastic waves $\lambda_{el}$ and the plate length $L_p$, so that the friction force can be approximated by a point force. This simplification not only brings analytic insights, but also has practical relevancy. For instance, in the existing experiment that the gold micro-plate azimuthally moves around the micro-fiber~\cite{Lu2019}, where the contact length in the azimuthal direction is negligible, as is shown the left panel of Fig.~\ref{fig::simple}, this single-friction-point model (SFPM) captures the real physics. Admittedly, there are cases where the PFFM is defective, e.g., when studying that a flat micro-plate translates along the axial direction of a micro-fiber, as is shown in the right panel of Fig.~\ref{fig::simple}. Then, the spatial distribution of the friction force is expected to play a role, for which the simplest extension is to consider two friction points, as shown in the right panel of Fig.~\ref{fig::simple} (see more discussions in Sec. VI).

{\it Derivations of SFPM.} --- Consider that a point friction force, denoted by $F_f$, is applied to the plate at $x=x_0$. The $F_f$, similar to the temperature variation $\delta T$, also induces the deformation of the plate. In the nanosecond regime and at micro-scales, the friction-induced elastic displacements also have to be evaluated from the elastic wave equation. Specially, the induced dominant $x$-component elastic displacement, $u_x^f$, is formulated as
\begin{subequations}
\begin{align}
u_x^f (x,t)=\int_{-\infty}^t G_{el} (x,x_0;t-t') F_f (t' )dt',
\label{eq:uxGreena}
\end{align}
where $G_{el} (x,x_0;t-t')$, the Green’s function of the linear elastic equation (see Supplementary Technical Note for its definition~\cite{SeeSupplementalMaterial}), can be expanded by the fundamental longitudinal elastic modes [i.e., Eq.\eqref{eq:elasK}], and is given by:
\begin{align}
G_{el} (x,x_0;t-t')\simeq \frac{t_R}{4M_p}  \sum_{s=0}^3\sum_{n=0}^\infty H(t-t'-t_s-nt_R ).
\label{eq:uxGreenb}
\end{align}
\end{subequations}
Here, $M_p$ denotes the mass of the 2D plate; $t_s$’s ($s$=0,1,2,3) denote the travel time of the elastic waves from $x_0$ to the point $x$ four times in one round trip, with $t_0=|x-x_0 |/v_L$, $t_1=t_R-|x-x_0 |/v_L$, $t_2=|L_p-x-x_0 |/v_L$ and $t_3=|L_p+x+x_0|/v_L$; $H(t)$ is the Heaviside step function with $H(t)=1$ for $t\ge0$ and $H(t)=0$ otherwise; $n=0,1,2,3,\cdots$ label the round trips of the elastic waves bouncing back and forth inside the plate. The validity of Eq.~\eqref{eq:uxGreenb} is numerically confirmed in Supplementary Fig. S2~\cite{SeeSupplementalMaterial}. The $y$-dependencies in Eq.~\eqref{eq:uxGreena} and~\eqref{eq:uxGreenb} are dropped since we here consider thin micro-plates with thickness significantly smaller than micro-meters and the elastic displacements are approximately uniform in the thickness direction, as shall be validated numerically in Fig.~\ref{fig::singleF}B2 and Fig.~\ref{fig::wg}D2.

Interestingly, Eq.~\eqref{eq:uxGreena} can be reformulated to our familiar Newtonian second law when $F_f$ is a slowly varying function in $t_R$. In this case, the round-trip series can be approximate by integrals over time, so that there is $u_x^f=\int_{-\infty}^t\int_{-\infty}^{t'} F_f (t'' )/M_p dt'' dt'$, which is just the integral form of Newton’s second law.

The friction force $F_f$ depends on the elastic displacement induced by the temperature change $\delta T$, denoted by $u_x^{th}$ [see Eq.~\eqref{eq:elaswave}] and also on itself at the previous time. To reveal this dependence, we decompose $u_x^f$ into the direct propagation term $u_{x;0}^f$ and the remaining reflection term $u_{x;R}^f$: $u_{x;0}^f=t_R/(4M_p)  \int_{-\infty}^t H(t-t'-t_0)  F_f (t' )dt'$  and $u_{x;R}^f=u_x^f-u_{x;0}^f$. The total $x$-component elastic displacement $u_x$, including both thermal and friction contributions, is $u_x=u_{x;0}^f+u_{x;R}^f+u_x^{th}$. Next, to eliminate the integrals in $u_x^f$, we employ the time derivative to $u_x$ and introduce the deformation velocity $v_x\equiv\partial u_x/\partial t$. On the friction point ($x=x_0$), we have that
\begin{subequations}
\begin{align}
v_x (x_0;t)=v_{x;0}^f (x_0;t)+v_{x;R}^f(x_0;t)+v_x^{th}(x_0;t), \label{eq:vxa}
\end{align}
with
\begin{align}
v_{x;0}^f(x_0;t)=\frac{t_R}{4M_p} F_f (t), \label{eq:vxb}
\end{align}                                                                                                                                         and $v_{x;R}^f(x_0;t)=t_R/(2M_p) \sum_{n=1}^\infty F_f (t-nt_R ) +t_R/(4M_p)  \sum_{s=2}^3 \sum_{n=0}^\infty F_f (t-nt_R-t_s )dt'$ depending on $F_f$ at the previous time.
\end{subequations}

In Eq.~\eqref{eq:vxa}, the instantaneous effect of the friction force on the elastic displacement is to induce $v_{x;0}^f$ to cancel with $v_{x;R}^f+v_x^{th}$, thus minimizing the magnitude of $v_{x;0}$. Specifically, when the friction point is still ($v_x=0$), we have $F_f=-4M_p (v_{x;R}^f+v_x^{th})/t_R$, which gives the expression of the static friction force. Besides, it is known that the static friction force is bounded by $|F_f |\le F_f^s$, where $F_f^s$ is the sliding resistance. When this bounding condition is broken, it means that the friction point slides, and, then, the dynamic friction force $F_f=-{\rm sgn}(v_x)F_f^d$ exerts on the plate, where  ${\rm sgn}(v_x )=\pm 1$ for $v_x>0$ and $v_x<0$, respectively. Note that the dynamic friction force $F_f^d$ is generally different from $F_f^s$. Nevertheless, for convenience of our analysis, we here simply put $F_f^d=F_f^s$.

Given the above physical arguments, a strategy to determine the friction force is formulated as follows:
\begin{subequations}
\begin{align}
F_f =- 4M_p (v_{x;R}^f +v_x^{th})/t_R \label{eq:Ffa}
\end{align}
if $|F_f|<F_f^s$, i.e., $|v_{x;R}^f (t)+v_x^{th} (t)|< F_f^s t_R/ 4M_p$;
\begin{align}
F_f =-{\rm sgn}[v_{x;R}^f+v_x^{th} ]F_f^s. \label{eq:Ffb}
\end{align}
otherwise.
\end{subequations}

\section{Effects of friction on thermal deformation: insights from Born-approximation of SFPM}

{\it Born approximation.} --- Equations (3)-(5) self-consistently define the SFPM, which can be immediately applied for numerical simulations. However, their use in analytic analysis is still hindered by the fact that the friction force at time $t$ correlates with itself at the previous time. To remove this difficulty, we assume $|v_{x;R}^f |\ll |v_x^{th}|$, and, accordingly, neglect $v_{x;R}^f$ in Eqs. (5). This approximation spiritually resembles the so-called Born approximation in the quantum wave scattering theory~\cite{Born1926}, where the leading-order wave scattering due to the incident waves (which, in our case,
is parameterized by $u_{x;0}^f$) is considered to be dominant, while the higher-order scattering terms (which here represent the elastic waves induced by the friction force at the previous time, $u_{x;R}^f$) are dropped. In the heating period when the thermal deformation significantly dominates over the friction, this approximation is proper. However, in the cooling period when the thermal deformation mitigates, $v_x^{th}$ and $v_{x;R}^f$ could become comparable, making the approximation invalid. Nevertheless, this invalidity should not bring serious effect on our qualitative understanding of the deformation dynamics, as long as the deformation mainly takes place in the intense-thermal-deformation period, which is particularly true with the use of short nanosecond laser pulses. Bearing these facts in mind, we proceed to derive the consequences of the Born approximation, which shall be supported by the full-numerical results.

Under the Born approximation, the strategy to determine the friction force, formulated in Eqs.~\eqref{eq:Ffa} and~\eqref{eq:Ffb}, could be simplified to: $F_f \simeq-4M_p v_x^{th}/t_R$, when $|v_x^{th} |<v^s$ with $v_x=0$; $F_f (t)\simeq-{\rm sgn}[v_x^{\rm th} (t)]F_f^s$ when $|v_x^{th}|>v^s$ with $v_x\simeq v_x^{th}-{\rm sgn}[v_x^{th} (t)] v^s$. Here, the threshold velocity $v^s$, which defines the minimum magnitude of the thermal deformation velocity required to enable the motion of the friction point,is given by
\begin{subequations}
\begin{align}
v^s=\frac{F_f^s t_R}{4M_p}.
\label{eq:vxs}
\end{align}
And the sliding distance of the friction point is
\begin{align}
u_x (t)\simeq\int_{-\infty}^t (v_x^{th} (t' )-{\rm sgn}[v_x^{th} (t' )] v^s )H(|v_x^{th} (t' )|-v^s)dt',
\label{eq:uxapprox}
\end{align}
where $H(x)$ is the Heaviside step function that is defined below Eq.~\eqref{eq:uxGreenb}.
\end{subequations}

{\it Implications from Eqs. (6)}. —-- (1) To realize a large sliding distance, it is necessary to enlarge the asymmetry in the heating and cooling timescales. Otherwise, $v_x^{th}$  in the heating and cooling periods, intend to cancel each other in Eq.~\eqref{eq:uxapprox}, thus reducing the net sliding distance. This requires either $t_{heat}\gg t_{cool}$ or $t_{heat}\ll t_{cool}$. The former condition $t_{heat}\gg t_{cool}$ demands the use of a pulse with the rising time much longer than $t_{cool}$. At the same time, the cooling should be fast enough to make thermal deformation velocity $v_x^{th}$ exceed $v^s$. On the other hand, the condition $t_{heat} \ll t_{cool}$ can be much more easily met by using a short light pulse with $t_w\ll t_{cool}$ (note that the heating time scale $t_{heat}$ is determined by $t_w$), which we focus on in the present article.

Having the timescale asymmetry between the heating and cooling periods, the sliding distance, approximated with Eq.~\eqref{eq:uxapprox}, shall be mainly contributed from the dominant thermal phase which has a larger magnitude of $v_x^{th}$. The stabilized sliding distance as $t\to \infty$ is then expected to point in the same direction as the thermal deformation in the dominant thermal phase.

\begin{figure*}[ht!]
\includegraphics[width=0.82\textwidth]{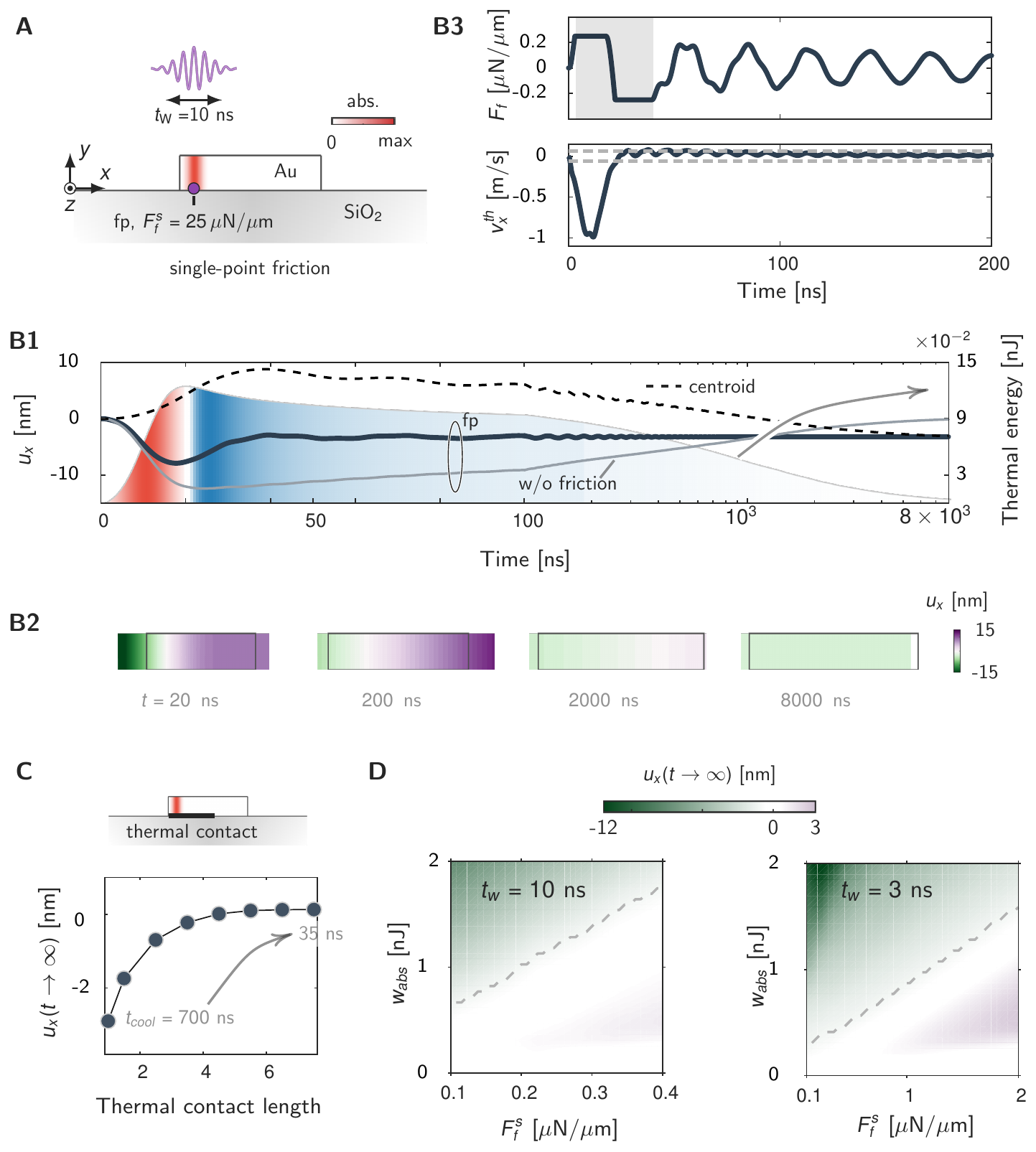}
\caption{{\bf Sliding dynamics with a single friction point driven by a nanosecond light pulse}. {\bf A.} Sketch of the investigated problem, which adds a point friction at 1 $\mu$m from the left edge of the plate to the “configuration 2” system studied in Fig.~\ref{fig::defwofric}. {\bf B1.} Temporal evolutions of thermal energy (shaded area), and $x$-component elastic displacements of the friction point (solid lines) and the plate centroid (dashed line). {\bf B2.} Profiles of $x$-component elastic displacements at different time $t$=20,200,2000,8000 ns. {\bf B3.} Temporal evolutions of $x$-component friction force (upper panel) and thermal deformation velocity (lower panel) of the friction point. The gray area in the upper panel marks the sliding period, while the dashed lines in the lower panel mark the predicted sliding threshold velocity $v^s$. In {\bf B1}-{\bf B3}, the injected light pulse has $t_w=10\, \rm, ns$ (temporal width) and $W_{abs}=0.2\, \rm nJ/\mu m$. The friction point has  $F_f^s=0.25\, \mu\rm N/\mu m$. {\bf C.} Net sliding distance as a function of the thermal contact length. As the thermal contact length increases from 1 $\mu$m (that is in {\bf B1-B3}) to 7.5 $\mu$m, the cooling time decreases from 700 ns to 35 ns. D. Net sliding distances as functions of the optical absorption energy $W_{abs}$ and friction resistance $F_f^s$ for $t_w=10\, \rm ns$ (left panel) and  $t_w=3\,\rm ns$ (right panel). The dashed lines trace the sliding distance of -1 nm in the $W_{abs}-F_f^s$ parameter space. The other simulation settings in {\bf D} are as the same as in {\bf B1-B3}.
}
\label{fig::singleF}
\end{figure*}

(2) To realize sliding against a larger friction, it is better to use shorter light pulses (with smaller $t_w$) than longer ones.  This is because that, to enable the motion of the friction point, the magnitude of the thermal deformation velocity $v_x^{th}$ should exceed the threshold velocity $v^s$, otherwise, the friction point remains still. This prerequisite calls for a rapid thermal deformation, which mainly occurs in the heating period due to our consideration $t_{heat}\ll t_{cool}$ (see the above discussions about the first implication). Specifically, considering the existing experiments summarized in Fig.~\ref{fig::table1}, where the typical size of the gold plate is about $\sim$10 $\mu$m $\times$ 10 $\mu$m$ \times$ 50 nm, indicating that $M_p\sim 10^{-13}-10^{-12}\,  \rm kg$ and $t_R\sim 10\, \rm ns$; moreover, it has been measured that $F_f^s\sim \mu \rm N$~\cite{Lu2019,Gu2021}. Thus, from Eq.~\eqref{eq:vxs}, the threshold velocity $v_x^s$ is inferred to be in the order of cm/s. As is shown in Fig.~\ref{fig::defwofric}, with the use of the nanosecond pulsed light, $u_x^{th}$ rapidly varies within the heating period, when $v_x^{th}$ can easily approach the order of m/s, significantly exceeding $v^s$. On the contrary, if continuous light is used, the related thermal deformation is gentle, and the nano-motion disappears, which has been validated experimentally~\cite{Lu2019,Lyu2022}.
Furthermore, as $t_w$ decreases, the thermal deformation velocity $v_x^{th}$ (in the heating period) increases, thus making easier exceed the threshold velocity $v^s$. Below, our numerical studies shall mainly focus on the use of nanosecond pulsed light echoing the same choice in the existing experiments~\cite{Lu2019,Lyu2022,Tang2021,Gu2021}. Nevertheless, as implied by our theoretical analysis, the use of shorter short light pulse, such as picosecond and femtosecond laser pulses, is also possible, which deserves further experimental explorations.

{\it Numerical validations.} ---  We numerically validate the aforementioned implications by revisiting the numerical case “configuration 2” studied in Fig.~\ref{fig::defwofric}.  A point friction force with $F_f^s=0.25\, {\mu}\rm N/\mu m$ (note that the unit of $F_f^s$ is $\mu\rm N/\mu m$ due to the 2D nature of the plate) is introduced at the center of the thermal contact, a light pulse with temporal width $t_W=10\,\rm ns$ and absorption energy $W_{abs}=0.2\,\rm nJ/\mu m$ is used, as is sketched in Fig.~\ref{fig::singleF}A (see the figure caption for more details). The numerical computations are implemented by using Eqs. (3)-(5) to exactly include the friction contributions (see Supplementary Technical Note for details of numerical implementations~\cite{SeeSupplementalMaterial}).

The shaded region in Fig.~\ref{fig::singleF}B1 illustrates the temporal evolution of the thermal energy, which features the asymmetry in the heating and cooling timescales, i.e., fast heating and slow cooling. The dark solid line in Fig.~\ref{fig::singleF}B1 plots the temporal evolution of the $x$-component elastic displacement of the friction point, $u_x$. It shows that the friction point mainly slides in the heating period, when the thermal deformation is intense, while mildly moving in the cooling period, when the thermal deformation is gentle. As a result, as $t\to\infty$, the plate accumulates a negative sliding distance, consistent with the observation from the Born approximation. On the contrary, without the friction, the friction point returns back to its original position by the slow thermal contraction, and the net sliding distance is zero, as is illustrated by the gray solid line in Fig.~\ref{fig::singleF}B1.

For a better visualization of the sliding dynamics, we plot the friction force $F_f$ and the thermal deformation velocity $v_x^{th}$ as functions of time in Fig.~\ref{fig::singleF}B3. The sliding period is outlined by the shaded regions in the upper panel of Fig.~\ref{fig::singleF}B3, where the magnitude of the friction force equals to $F_f^s=0.25\, \mu\rm N/\mu m$ .  Particularly, in the heating period, the sliding is along the $-\hat x$ direction and the friction force is thus positive. Notably, it appears when $|v_x^{th} |>v^s$  (see the lower panel of Fig.~\ref{fig::singleF}B3 where the dashed lines mark $|v_x^{th} |=v^s$), as predicted from the Born approximation. On the contrary, in the early cooling period, the sliding is along the $\hat x$ direction and the friction force becomes negative. In this period, due to the negligible magnitude of $v_x^{th}$, the contribution from the friction force at the previous time [characterized by $v_{x;R}^f$ in Eqs. (4)] becomes relatively important and, thus, the Born approximation is defective. As a result, the sliding condition is no longer described by $|v_x^{th}|>v^s$.

We also evaluate the $x-$component centroid displacement of the plate (denoted by $<u_x>$; see dashed line Fig.~\ref{fig::singleF}B1) with Newton's second law $M_p\partial ^2<u_x>/\partial t ^2 = F_f$. It is initially in the opposite direction to the displacement of the friction point, and then gradually approaches the latter. This can be understood by examining the temporal evolution of the friction force, as is shown in the upper panel of Fig.~\ref{fig::singleF}B3. Further, as is plotted in Fig.~\ref{fig::singleF}B2, the profiles of the $x$-component elastic displacements of the plate at $t$=20,200,,2000,8000 ns show that the left and right sides of the plate are initially stretched in opposite directions by the thermal deformation, and then, the two sides gradually crawl toward the friction point that is anchored by the friction force. Finally, the entire plate gets the same displacement.

We next illustrate the effects of the thermal asymmetry by varying the cooling time $t_{cool}$, while maintaining other parameters unchanged. The cooling time is modified by changing the thermal contact length between the plate and the substrate, as illustrated in the inset of Fig.~\ref{fig::singleF}C. Initially, this length (as in Fig.~\ref{fig::defwofric}B) is 1 $\mu$m  that leads to $t_{cool}\simeq 700 ns$. Increasing it to 7.5 $\mu$m, $t_{cool}$ is reduced to 35 ns, close to $t_w=10\,\rm ns$.  As is discussed above, the intensified thermal cooling, which reduces the thermal asymmetry, drags the friction point more back to its initial position in the cooling period, thus decreasing the net sliding distance, as is confirmed in Fig.~\ref{fig::singleF}C.

Further, we study the benefits of using shorter light pulses to overcome greater friction. By varying the injected optical absorption energy $W_{abs}$ and the sliding resistance $F_f^s$ (the other parameters are set as the same as in Figs.~\ref{fig::singleF}B1-B3), the net sliding distances for $t_w=3,10\,\rm ns$ are plot in Fig.~\ref{fig::singleF}D. The results show that, under the injection of a small amount of nanojoule absorption energy, the shorter pulse with $t_w= 3\,\rm ns$ can enable the sliding of the plate under larger friction resistance than the longer pulse with $t_w= 10\,\rm ns$. This fact can be more clearly appreciated by observing the dashed lines in Fig.~\ref{fig::singleF}D, which mark an exemplified sliding distance of -1 nm and show that the same sliding distance is achieved under a larger friction with the use of a shorter pulse.  Moreover, we note that, in the most area of the simulated $W_{abs}-  F_f^s$ parameter space, the sliding distance is negative (green color). This agrees with the insight from the Born approximation that the plate slides along the same direction as the thermal deformation of the friction point in the heating period. However, there exist also small purple regions, showing the positive sign of the sliding distance. This “contradiction” manifests the failure of the Born approximation, and occurs when the friction-induced elastic waves become comparable to the thermal-induced ones.

\section{Beyond single-friction-point model: effects of friction distribution}

\begin{figure*}[ht!]
\includegraphics[width=0.8\textwidth]{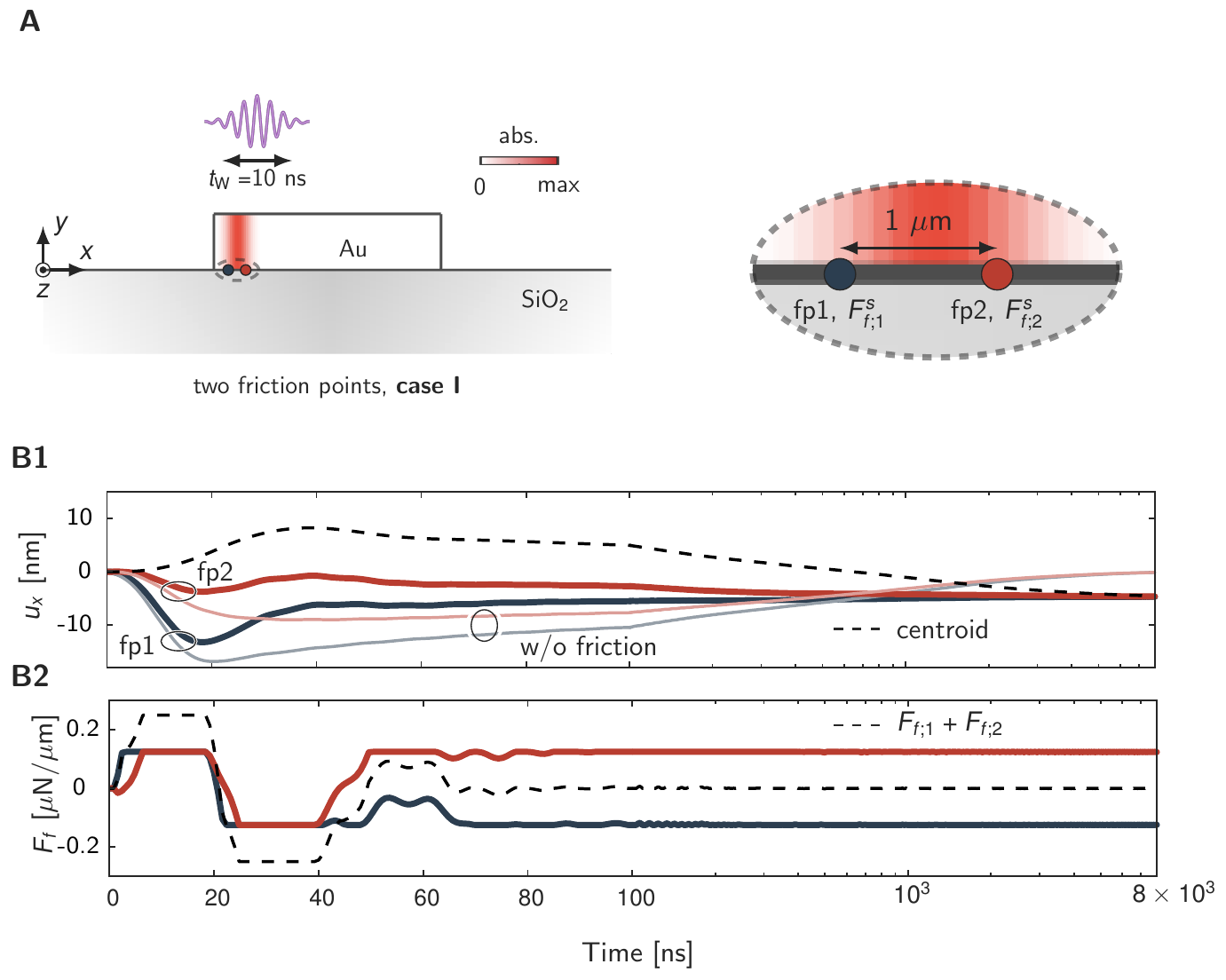}
\caption{{\bf Sliding dynamics with two close friction points (case I) driven by a nanosecond light pulse.} {\bf A.} Sketch of the investigated problem. The right enlarged view shows two friction points with 1 $\mu$m separated distance, and the distance from the left friction point to the left edge of the plate is 0.5 $\mu$m. {\bf B1.} Temporal evolutions of $x$-component elastic displacements of two friction points with (dark solid lines) and without (gray solid lines) friction force, and of the centroid displacement (dashed line). {\bf B2.} Temporal evolutions of $x$-component friction forces of two friction points (solid lines) and their summation (dashed line). In {\bf B1- B2}, the friction resistances of two friction points are $F_{f;1}^s=F_{f;2}^s=0.125 \,\mu\rm N/\mu m$. The injected light pulse has $t_W=10\,\rm ns$ and $W_{ abs}=0.2\,\rm nJ/\mu m$. The other simulation parameters take the same settings as the “configuration 2” in Fig.~\ref{fig::defwofric}.
}
\label{fig::douleF1}
\end{figure*}

As the friction dimension along the sliding direction increases, the spatial distribution of the friction force, beyond the single-point-friction consideration, needs to be taken into account. To illustrate this effect, the simplest extension is to introduce two friction points and study their joint effects on the thermal deformation. Practically, referring to a flat or partially curved plate on a substrate, as is sketched in the right panel of Fig.~\ref{fig::simple}, this extension could be regarded as a crude approximation to the real continuous contact by dividing the contact region in half and concentrating the friction force in each half into a point force.

{\it Case I.} --- First, to establish a link with the SFPM, we split the single friction point in Figs.~\ref{fig::singleF} into two with a small separating distance of 1 $\mu$m. The sliding resistances of two points are set to be  $F_{f;1}^s=F_{f;2}^s=0.125 \,\mu\rm N/\mu m$, as is sketched in Fig.~\ref{fig::douleF1}A. In this case, the distance between the two points is significantly smaller than the plate length, so that their thermal deformation properties are expected to be similar. Moreover, their distance is also smaller than the characteristic elastic wavelength $\lambda_{el}=t_w v_L\simeq 20\, \rm \mu m$ ($t_w=10\,\rm ns$), implying that the retardation effects of the elastic waves between two points are negligible. Given these facts, we intuitively expect that the two friction points considered here might lead to the sliding dynamics close to the single-friction-point case as illustrated in Figs.~\ref{fig::singleF}. The present numerical study is to confirm this intuition, and, moreover, to reveal apparent features that are absent in the SFPM.

The temporal evolutions of the $x$-component elastic displacements of two friction points (see thicker solid lines in Fig.~\ref{fig::douleF1}B1; see the figure caption for the simulation details) exhibit features similar to the single-friction-point case. More precisely, the friction points slide rapidly in the heating period, and, then, gradually ceases their motions and converge to the centroid displacement (dashed line) in the cooling period.

However, the friction-force evolutions, as shown in Fig.~\ref{fig::douleF1}B2, show a feature that is absent in the SFPM. Specifically, after an initial slip ($t<40\,\rm ns$) during which the friction forces of two points show the same direction, they turn into opposite directions. The friction force of the left point (dark blue line) is in the $-\hat x$ direction, while that of the right point (dark red line) is in the $\hat x$ direction. The friction forces thus stretch the part of the plate between the two friction points. At the same time, the thermal contraction in the cooling period slowly drags the two points to approach each other. Finally, with the end of the thermal cooling, the friction forces remain at the same opposite non-zero values equal to the sliding resistance, while their summation vanishing (dashed line).

\begin{figure*}[ht!]
\includegraphics[width=0.8\textwidth]{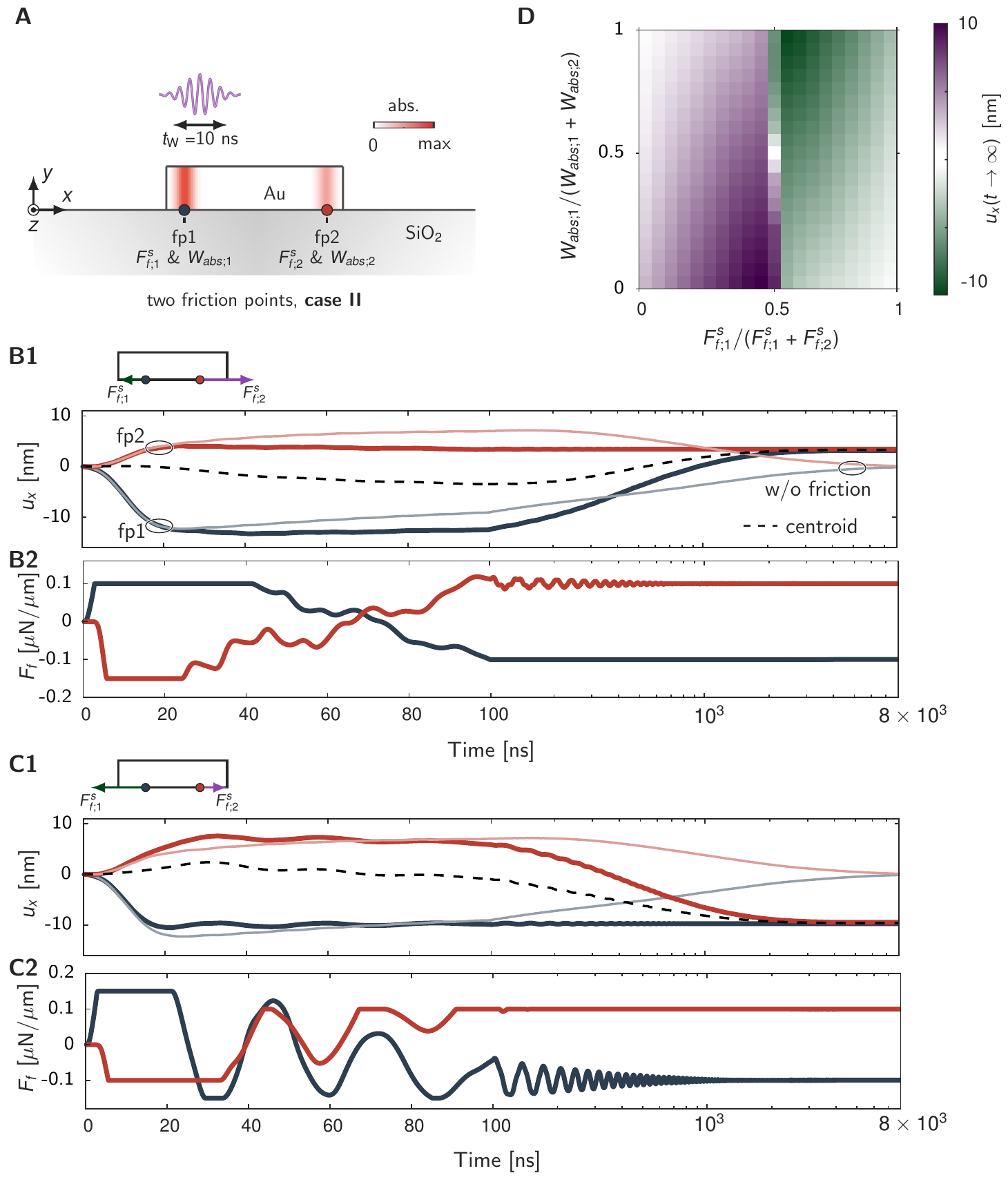}
\caption{{\bf Sliding dynamics with two distant friction points (case II) driven by a nanosecond light pulse.} {\bf A.} Sketch of the investigated problem. Two friction points are introduced with 1 $\mu$m distance to the left and right edges of the plate. The light absorption is distributed around two friction points with localized energy denoted by $W_{abs;1}$ and $W_{abs;2}$. {\bf B1.} Temporal evolutions of $x$-component elastic displacements of two friction points with (dark solid lines) and without (gray solid lines) friction force, and of the centroid displacement (dashed line). {\bf B2.} Temporal evolutions of $x$-component friction forces of two friction points (solid lines) and their summation (dashed line). In {\bf B1-B2}, the friction resistances of two friction points are $F_{f;1}^s=0.1\, \rm \mu N/\mu m$  and $F_{f;2}^s=0.15\, \mu\rm N/\mu m$. The injected light pulse has $t_W=10\,\rm ns$ and $W_{abs}=0.2 \, \rm nJ/\mu m$. The other simulation parameters take the same settings as the “configuration 2” in Fig.~\ref{fig::defwofric}. {\bf C1.-C2.} Same as {\bf B1-B2}, but with $F_{f;1}^s=0.15\,\rm \mu N/\mu m$  and $F_{f;2}^s=0.1\, \mu \rm N/\mu m$. {\bf D.} Net sliding distance as functions of $W_{abs;1}/(W_{abs;1}+W_{abs;2})$ and  $F_{f;1}^s/(F_{f;1}^s+F_{f;2}^s)$ with  $W_{abs;1}+W_{abs;2}=0.2\, \rm nJ/\mu m$ and $F_{f;1}^s+F_{f;2}^s=0.25\,\rm \mu N/\mu m$.
}
\label{fig::douleF2}
\end{figure*}

{\it Case II.} --- We then increase the distance between two friction points by setting that they locate with 1 $\mu$m distance to the left and right edges of the plate, respectively, as is sketched in Fig.~\ref{fig::douleF2}A. In this case, the distance between two points (8 $\mu$m) are comparable to the plate length, so that they can be no longer approximated as the single point as in Figs.~\ref{fig::douleF1}. Particularly, the thermal deformations of such two points are in the opposite directions. This naturally brings one question that cannot be simply answered by referring to the insights of the SFPM: in which direction the plate should slide?

Clearly, two factors, the spatial distributions of (1) the friction forces and (2) the light absorption, are possibly involved in determining the sliding direction. To illustrate the second factor, we artificially set the absorption profile by rearranging them around two friction points with the localized energy denoted by $W_{abs;1}$ (left point) and $W_{abs;2}$ (right point), as is shown in Fig.~\ref{fig::douleF2}A. Bearing these factors in mind, we start our investigations by intentionally setting that the two friction points have different sliding resistances with $F_{f;1}^s=0.1\, \mu\rm N/\mu m$ (left point) and $F_{f;2}^s=0.15\, \mu\rm N/\mu m$,  (right point), for reasons that will come clear later; the absorption profile is set with $W_{abs;1}=0.2\,\rm nJ/\mu m$ and $W_{abs;2}=0$. The other simulation parameters are as the same as in Figs.~\ref{fig::douleF1}B1-B3.

The solid thicker lines in Fig.~\ref{fig::douleF2}B1 illustrate the temporal evolutions of the $x$-component elastic displacements of the left and right friction points, while the sold thinner lines, the same evolutions but without the friction, are also included for highlighting the roles of the friction forces. We observe that, during the heating period, the two points slide intensely in the opposite directions. Then, in the early cooling period (e.g., $20\,{\rm ns} <t<100\,{\rm ns}$), their motions are significantly suppressed with negligible displacements. Then, over time ($t>100\,\rm ns$), the displacement of the left friction point with a smaller sliding resistance gradually approaches that of the right friction point with a larger sliding resistance, while the right friction point remains almost still. As a result, the net sliding is towards the right direction, as is confirmed by the centroid displacement (dashed line shown in Fig.~\ref{fig::douleF2}B1).

We further examine the temporal evolutions of the friction forces in Fig.~\ref{fig::douleF2}B2. In the initial sliding phase (i.e., heating period), the friction forces of two points point in the directions opposite to the sliding directions with different magnitudes given by $F_{f;1}^s$ and $F_{f;2}^s$, respectively. Later, they gradually reverse their signs with the magnitudes approaching the smallest value of $F_{f;1}^s$ and $F_{f;2}^s$ (i.e., $F_{f;1}^s=0.1 \,\mu\rm N/\mu m$). Apparently, the friction force exerts on the right point is smaller than its sliding resistance, so that the right point is kept still by the static friction, while the left point moves towards the right side, as observed in Fig.~\ref{fig::douleF2}B1. Finally, in the stabilized state ($t\to\infty$), the friction forces of the left and right sides have the opposite signs and the same magnitudes of  $F_{f;1}^s$ (i.e., 0.1$\mu$N/$\mu$m), which should lead to a small tensile stress between two points, similarly as in the case I shown in Fig.~\ref{fig::douleF1}.

The dynamic evolutions of the displacements of two friction points physically resemble the well-known game, tug-of-war. In the first stage, as the game starts, the competing teams stretch the rope and slide on the ground, which just mimics the sliding of two friction points in the heating period enabled by the thermal deformation.  Then, the two teams adjust their body postures for larger friction against the sliding, and reach a certain static balance in a short period of time, which corresponds to the early cooling period shown in Fig.~\ref{fig::douleF2}B1. Eventually, the team subjected to a lager friction (e.g., due to heavier weight or wearing shoes with firmer grip) shall drag the other team toward them and win the game, which intuitively interprets that the left friction point with a smaller sliding resistance gradually slides towards the side with a larger sliding resistance.

Next, we reverse the values of $F_{f;1}^s$ and $F_{f;2}^s$ with $F_{f;1}^s=0.15\,\rm \mu N/\mu m$ and $F_{f;2}^s=0.1\,\rm \mu N/\mu m$, so that the left friction point now has a larger sliding resistance than the right point. The simulated the temporal evolutions of the elastic displacements and the friction forces are plotted in Figs.~\ref{fig::douleF2}C1 and \ref{fig::douleF2}C2, respectively. They show similar features as shown in Figs.~\ref{fig::douleF2}B1 and \ref{fig::douleF2}B2, except that the sliding direction now reverses to the $-\hat x$ direction due to that the left point with a larger sliding resistance drags the right point towards it.

The comparisons between Figs.~\ref{fig::douleF2}B1-B2 and Figs.~\ref{fig::douleF2}C1-C2 suggest that the sliding direction is mainly determined by the friction distribution. The distribution of the light absorption, which mainly determines the strength of the thermal deformation in the early heating period, seems not to be decisive. To conclusively evidence this observation, we compute the net displacement of the plate by varying the distribution of the absorption profile, i.e., $W_{abs;1}/(W_{abs;1}+W_{abs;2})$ ($W_{abs;1}+W_{abs;2}=0.2\, \rm nJ/\mu m$), and, the friction distribution, i.e., $F_{f;1}^s/(F_{f;1}^s+F_{f;2}^s)$ ($F_{f;1}^s+F_{f;2}^s=0.25\,\rm \mu N/\mu m$), and plot the results are in Figs.~\ref{fig::douleF2}D. We see that the sliding direction (the sign of the sliding distance) is indeed determined by the relation between $F_{f;1}^s$ and $F_{f;2}^s$, and the absorption profile, parameterized by $W_{abs;1}/(W_{abs;1}+W_{abs;2})$, is almost insignificant. More precisely, when $F_{f;1}^s>  F_{f;2}^s$, i.e., the left friction dominates over the right one, the sliding direction is towards the left, and vice versa. Only in the case that $F_{f;1}^s$ and $F_{f;2}^s$ approach each other, that is, $F_{f;1}^s/(F_{f;1}^s+F_{f;2}^s)$  is close to 0.5, the sliding direction becomes dependent on the absorption distribution, which is towards the side with a lager absorption energy. Thus, these results suggest that the sliding direction is mainly determined by the friction distribution.

\section{Perspectives: a practical proposal for on-chip integration}

Up to now, all the existing experimental demonstrations of the nano-stepping motion driven by light-induced elastic waves are restricted in the microfiber-based systems (see Fig.~\ref{fig::table1}). Our insights developed above can be directly applied to better interpret these experimental observations. Nevertheless, instead of going into details of the existing experiments, we here ask an unexplored question: whether the same technique can be translated onto micro- or nano-waveguide systems, or even on planar substrates. We consider that the attempts in this direction are meaningful, yet challenging in the following aspects.

First, it is known that, in a microfiber-based system, the contact between the micro-fiber and the micro-object is confined in a narrow line-shaped region, which results in a friction force of a few $\mu$N~\cite{Lu2019,Gu2021}. Replacing the micro-fiber with the micro- and nano- waveguides that generally have planar roofs increases the contact area, and, thus, enhance the friction force. To put this in concrete terms, considering that a gold plate contacts with a micro-fiber with a diameter of 2 $\mu$m, the effective transverse contact length (perpendicular to the axial direction of the micro-fiber) due to the vdW force is estimated to be in the order of 10 nm (see Supplementary Fig. S1~\cite{SeeSupplementalMaterial}).  Then, using a nano waveguide with a transverse width about 0.5 $\mu$m, the effective contact area is increased at least more than tenfold, so that the friction force might reach tens of $\mu$N,  making the driving more challenging. Nevertheless, the insights from the previous physical modes provide some positive signs, suggesting that the use of short nanosecond pulses (e.g., with $t_w=3 \,\rm ns$, see Fig.~\ref{fig::singleF}D) is possible to enable the motion of a micro-object under the sliding resistance over 10 $\mu$N.  If this is realizable, one can even imagine that a  micro-object (subjected to a friction force of tens, or even hundreds of $\mu$N) on a planar substrate might even be manipulatable with the same principle.

Second, liberating the manipulations from the micro-fiber-based systems can lead to wider photonic applications. For instance, in micro- and nano-waveguide systems on optical chips, if a micro-object can be delivered into user-demanded locations precisely, it could be used to control light transmission in a single waveguide, or tune power coupling between nearby waveguides, by modifying photonic environments through its movable positions. Besides, if the vibration modes of the micro-object can be additionally exploited, their combination with position manipulations might bring other possibilities in opto-mechanics and practical applications, such as mobile optical modulators.

\begin{figure*}[ht!]
\includegraphics[width=0.8\textwidth]{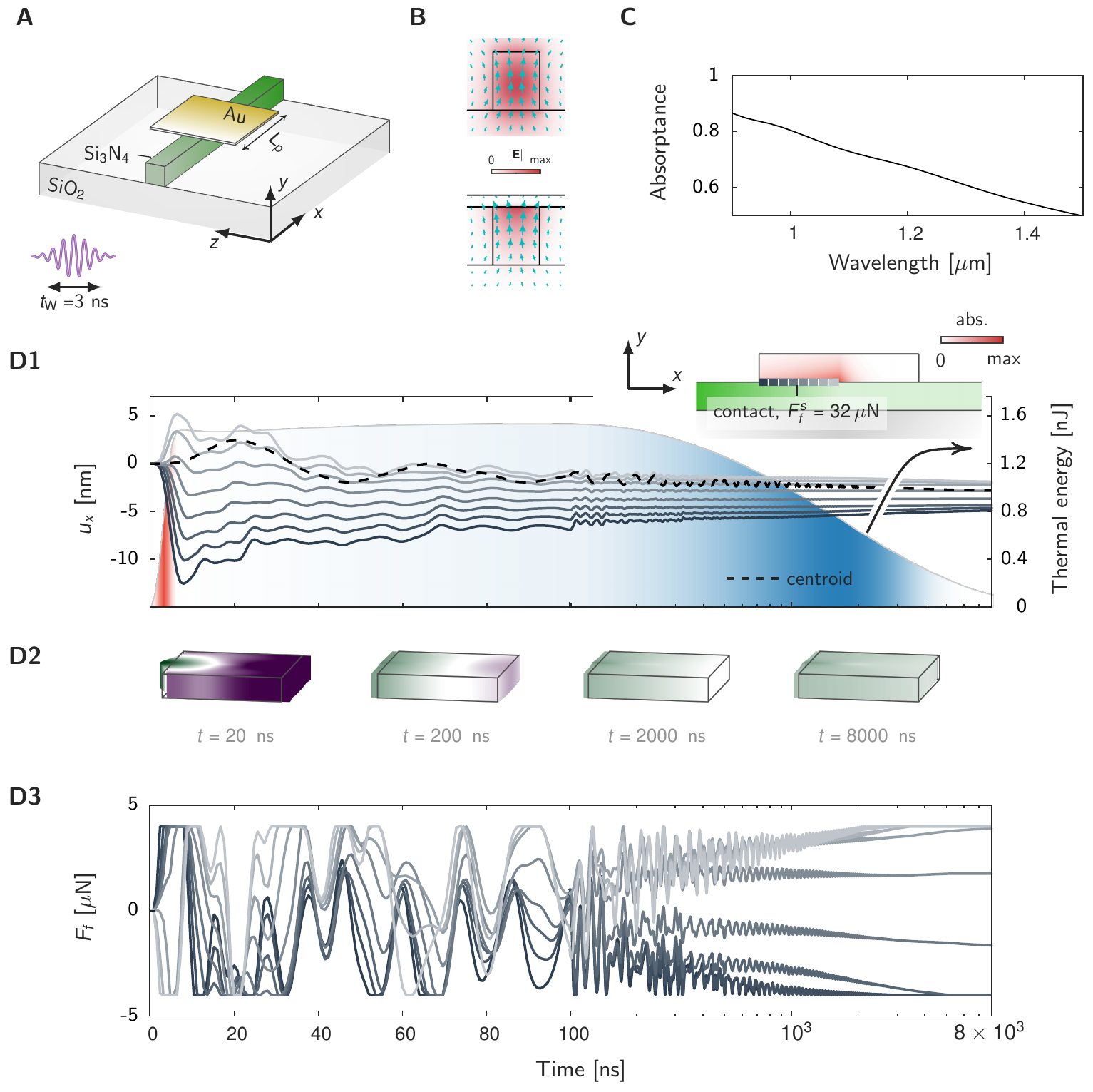}
\caption{ {\bf Sliding of a gold plate on a Si$_3$N$_4$ nano-waveguide driven by a nanosecond light pulse.} {\bf A.} Sketch of a gold nanoplate sitting above a Si$_3$N$_4$ nano-waveguide on a Si$_2$O$_2$ substrate. The Si$_3$N$_4$ nano-waveguide has a cross-sectional width of 400 nm and a height of 500 nm. The gold square plate has a slide length of 16 $\mu$m and thickness of 100 nm. {\bf B.} Modal profiles ($|\bf E|$ distributions) of fundamental TM mode of the Si$_3$N$_4$ nano-waveguide without (upper) and with (lower) the gold plate at light wavelength 1.03 $\mu$m. The arrows specify the directions of the in-plane E fields. {\bf C.} Absorptance spectrum of the hybrid gold-Si$_3$N$_4$ waveguide under incidence of the TM mode of the Si$_3$N$_4$ nano-waveguide. {\bf D1.} Temporal evolutions of thermal energy (shaded region) and $x$-component elastic displacements (solid lines). Inset: the left-half side of the plate contacting (both mechanically and thermally) with the nano-waveguide, leading to a sliding resistance of 32 $\mu$N and a cooling time over 1000 ns. {\bf D2.} Profiles of $x$-component elastic displacements at different time $t$=20,200,2000,8000 ns. {\bf D3.} Temporal evolutions of x-component friction forces. In {\bf D1} and {\bf D3}, the contact area is divided into eight equal segments, on which the averaged elastic displacements and the integrated friction forces are plotted. The temporal width and absorption energy of light pulses are $t_W$=10 ns and $W_{ abs}$=0.2 nJ/$\mu \rm m$, respectively.
}
\label{fig::wg}
\end{figure*}

Below, we provide realistic numerical simulations that positively evidence that a gold micro-plate can be driven by pulsed light on top of a nano-waveguide with friction force in the order of tens of $\mu$N.  These simulations also echo our developed theory, and show that the revealed physical insights can be directly applied for interpreting the physics in complex systems. The design of the nano-waveguide is sketched in Fig.~\ref{fig::wg}A. A Si$_3$N$_4$ nanowire sits on a SiO$_2$ substrate. Both Si$_3$N$_4$ and SiO$_2$ have higher specific heat capacity and much lower thermal conductivity than gold (see Supplementary Table 1~\cite{SeeSupplementalMaterial}), thus ensuring a slow cooling process. Concretely, consider a Si$_3$N$_4$ nanowire with width 400 nm and height 500 nm, which supports single-mode propagation of the transverse -magnetic (TM) light at near-infrared wavelengths. The upper panel of Fig.~\ref{fig::wg}B plots the modal profile of the TM waveguiding mode at 1.03 $\mu$m. Covering a gold plate on top of the Si$_3$N$_4$ nanowire, the modal energy shall be slightly redistributed and localized near the gold surface due to excitations of surface plasmon polaritons, as is shown in the lower panel of Fig.~\ref{fig::wg}B (gold plate thickness and side length: 100 nm and 16 $\mu$m, respectively). As the hybrid metal-dielectric mode propagates along the nanowire, the light gradually converts into heat due to the Ohmic losses of gold, and the overall absorptance is determined by the propagation length. Accordingly, as is shown Fig.~\ref{fig::wg}C, a high absorptance (above 50$\%$) is reached in a broad wavelength range from 0.9 $\mu$m to 1.5 $\mu$m.

To examine the actuation dynamics of the plate, we inject a nanosecond light pulse with $t_w=3\, \rm ns$ into the Si$_3$N$_4$ nanowire. The light pulse propagates along the $-\hat x$ direction (see the inset in Fig.~\ref{fig::wg}D1) and leads to a total energy  $W_{abs}=2\,\rm nJ$ absorbed by the plate. The plate is cut symmetrically by the nanowire in the $\hat z$ direction (see Fig.~\ref{fig::wg}A). Moreover, taking our previous experimental experience~\cite{Tang2021} that a thin gold plate is generally curved after its transferring process onto the waveguide, we assume that only the left half side of the plate contacts with the nanowire (c.f. the inset in Fig.~\ref{fig::wg}D1), so that the contact length in the $\hat x$ direction is 8 $\mu$m, indicating a contact area of  8 $\mu$m $\times$ 400 nm. This area is estimated to be about ten times larger than that of the same-sized plate placed on a micro-fiber with 2 $\mu$m diameter (see the discussions in the second paragraph in the same section). Moreover, it is known that the friction force is linearly proportional to the contact area. Therefore, from the measured friction force in the micro-fiber system that is about a few $\mu$N, we reasonably set that the friction force between the plate and the nanowire is $F_f^s=32 \,\mu\rm N$.

The shaded region in Fig.~\ref{fig::wg}D1 illustrates the temporal evolutions of the thermal energy stored in the plate. It shows that, after rapid inject of the optical energy by the short light pulse, the thermal energy slowly decays with the cooling time exceeding 1000 ns.  This confirms the intuition that the narrow Si$_3$N$_4$ nanowire with high heat capacity and low thermal conductivity can efficiently mitigate the heat conduction from the plate into the substrate.

The solid lines in Fig.~\ref{fig::wg}D1 show the temporal evolutions of the spatially averaged x-component elastic displacements of the eight segments that divide the contact area equally. They, in the early heating period, show noticeable differences mainly inherited from the spatial distribution of the thermal deformation. Then, they gradually approach each other, and the entire plate gains a net displacement in the $-\hat x$ direction (see the dashed line in Fig.~\ref{fig::wg}D1 for the centroid displacement).  This process can be better visualized in Fig.~\ref{fig::wg}D2 that shows the distributions of the x-component displacements at different time.

In the stabilized state ($t\to\infty$), the $x$-component displacement on the contact area shows an inhomogeneous distribution. This is similar to the simplified two-friction-points case that the displacements of the two friction points are slightly different with each other due to the “tug-of-war” like friction stretching. The solid lines in Fig.~\ref{fig::wg}D3 show the temporal evolutions of the friction force integrated in the divided segments of the contact area, and exhibit the features similar to the two-friction-points case. Initially, the left (darker lines) and right (grayer lines) sides on the contact area experience the friction forces pointing towards the right and left sides, respectively, to resist the sliding due to the thermal deformations. Then, after rather complicated evolutions when $t>100\,\rm ns$, the friction forces of the two sides reverse their signs and stretch the plate.

When replacing the Si$_3$N$_4$ waveguide with a micro-fiber with diameter of a few $\mu m$, the transverse contact length in the $z$ direction is reduced to tens of nanometers (see Supplementary Fig. S1~\cite{SeeSupplementalMaterial}). In this case, the sliding resistance is accordingly reduced. For instance, considering $F_f^s=3.2\,\mu\rm N$ that quantitatively mimics the contact between the gold plate and the micro-fiber~\cite{Tang2021}, the results in Supplementary Fig. S4~\cite{SeeSupplementalMaterial} demonstrate that the sliding shows similar features as in Fig. 8, thereby implying that the intense thermal deformation dwarfs the impact of the different sliding resistances in the two cases. Moreover, we numerically find that the sliding is allowable even when $F_f^s$ reaches to $80\,\mu \rm N$, see Supplementary Fig. S5~\cite{SeeSupplementalMaterial}.

To end this section, we summarize that the results in Fig.~\ref{fig::wg} encouragingly confirm that this elastic-wave-based manipulation principle is possible to be translated to micro/nano-scale optical waveguides on optical chips.

\section{Conclusions}

In this article, we comprehensively study the physical mechanism of nano-motion of micro-objects driven by elastic waves induced by nanosecond laser pulses. Focusing on interpreting roles of key optical-thermal-elastic quantities/parameters involved in the physical processes, including light pulse duration and energy, thermal heating and cooling time, sliding resistance and elastic waves, we develop a pedagogical single-friction-point model, which reveals that the use of short pulses and the thermal asymmetry in the heating and cooling timescales are two key factors to enable the motion against $\mu$N-friction. Further, we discuss the effects of the friction distribution beyond the single-friction-point consideration and show the tug-of-war effects due to the friction stretching. We envision that the studied manipulating principle can be translated to micro- and nano-waveguide systems on optical chips, and provide the numerical confirmations. Our theoretical results are expected to help future developments of optical manipulation on solid interfaces~\cite{Zheng2022b}.

{\bf Acknowledgements}---This project was supported by the National Natural Science Foundation of China (62275221).

\bibliography{References}

\end{document}